\documentclass[10pt]{article}
\def\D{\Delta}
\def\d{\delta}
\def\L{\Lambda}
\def\l{\lambda}
\def\S{\Sigma}
\def\G{\Gamma}
\def\g{\gamma}
\def\e{\epsilon}

\def\o{\omega}
\def\i{\iota}
\def\a{\alpha}
\def\b{\beta}

\def\dim{\textrm{dim}}
\def\det{\textrm{det}}
\newcommand{\be}{\begin{equation}}
\newcommand{\ee}{\end{equation}}
\newcommand{\bea}{\begin{eqnarray}}
\newcommand{\eea}{\end{eqnarray}}

\begin{document}

\begin{center}
\bf{NEW SPIN FOAM MODELS OF QUANTUM GRAVITY}\footnote{Talk given at the
Workshop on Quantum Gravity and Noncommutative Geometry, Lusofona University,
20 - 23 July 2004, Lisbon. Work supported by the FCT grant POCTI/MAT/45306/2002.}
\end{center}

\bigskip
\bigskip
\begin{center}
A. MIKOVI\'C
\end{center}

\begin{center}\textit{Departamento de Matem\'atica  \\
Universidade Lus\'ofona de Humanidades e Tecnologias\\
Av. do Campo Grande, 376, 1749-024 Lisbon, Portugal\\
E-mail: amikovic@ulusofona.pt}
\end{center}

\bigskip
\bigskip
\begin{quotation}
\small{We give a brief and a critical review of the Barret-Crane spin foam models of 
quantum gravity. Then we describe two new spin foam models which are obtained by direct quantization of General Relativity and do not have some of the drawbacks of the Barret-Crane models. These are the model of spin foam invariants for the embedded spin networks
in loop quantum gravity and the spin foam model based on the integration of the tetrads 
in the path integral for the Palatini action.}\end{quotation}

\bigskip
\bigskip
\noindent{\bf{1. Introduction}}

\bigskip
\noindent The spin foam models of quantum gravity represent a way to define
the path-integral for General Relativity in the Cartan formalism, i.e. insted of using the
four-metric $g_{\mu\nu}$ as the basic variable, one uses the tetrad one-forms
$e_{\mu}^a dx^{\mu}$ and the spin connection one-forms $\o_{\mu}^{ab}dx^{\mu}$. The Einstein-Hilbert action becomes the Palatini action
\be S=\int \e_{abcd} \,e^a \wedge e^b \wedge R^{cd} \quad,\ee
where $R_{ab}=d\o_{ab} + \o_a^c\wedge\o_{ca}$, so that one has to define the path-integral
\be Z = \int {\cal D} e\, {\cal D} \o e^{i\int \e_{abcd} e^a \wedge e^b \wedge R^{cd} }\quad.\ee

Notice that if one introduces a two-form
\be B_{ab}=\epsilon_{abcd} \,e^c \wedge e^d \quad,\label{bcon}\ee
then the Palatini action can be rewritten as the $SO(3,1)$ BF theory
action 
\be S=\int_M Tr (B\wedge F) \quad,\ee
where $F=dA + A\wedge A$ and $A=\o$. The BF action defines a topological theory,
so that in order to obtain Genral Relativity, one needs to impose the constraint 
(\ref{bcon}). Therefore one may try to define the GR path integral by using the BF theory
path integral and then constraining it, which was the strategy adopted by Barrett and Crane
\cite{bce,bc}.

The BF theory path integral can be defined as a sum over the irreducible representations (irreps)
of the BF theory Lie group $G$ of the amplitudes constructed by labeling the faces of the dual 2-complex of a triangulation of the manifold $M$ with these irreps \cite{bu,o}.  
One can arrive to this definition by starting from 
\bea Z &=& \int {\cal D}A\,{\cal D}B\,
\exp\left(i\int_M Tr(B\wedge F)\right)\nonumber\\ &=& \int \prod_l
dA_l \prod_\D dB_\D \exp\left(i\sum_{f} Tr(B_\D F_f
)\right)\quad,\eea where $l$ and $f$ are the edges and the faces
of the dual two-complex $\cal F$ for the simplical complex $T(M)$, while
$\D$ are the triangles of $T$. The variables $A_l$ and $B_\D$ are defined
as $\int_l A $ and $\int_\D B $ respectively, while $F_f = \int_f F$.

By performing the $B$ integrations
one obtains \be Z= \int \prod_l dA_l \prod_f \delta (F_f) \quad,\ee
which can be defined as \be Z= \int \prod_l dg_l \prod_f \delta
(g_f) \quad,\ee where $g_l = e^{A_l}$ and $g_f = \prod_{l\in\partial f} g_l$. By using
the well-known identity for the group delta function
\be \delta(g) = \sum_\L \textrm{dim}\,\L \,\chi_\L (g) \quad,\ee
where $\L$'s are the group irreps and
$\chi$'s are the corresponding characters,
one obtains
\be Z= \sum_{\L_f,\iota_l} \prod_f \dim\,\L_f \prod_v A_v
(\L_f,\iota_l) \quad,\label{tsfss}\ee
where $A_v$ is the vertex amplitide associated to the 4-simplex dual to
the vertex $v$. This amplitude is given
by the evaluation of the corresponding 4-simplex spin network, known as the
$15j$ symbol.

The sum (\ref{tsfss}) is called a spin foam state sum, because it is a sum of
the amplitudes for the two-complex $\cal F$ labelled with spins (irreps), i.e. a spin foam \cite{b}. However, the expression (\ref{tsfss}) is generically divergent, and this requires a regularization. A topologically invariant regularization is to replace the irreps
of $G$ with the irreps of the quantum group $G_q$, where $q$ is a root of unity. The form of the state sum stays the same, but now $\dim \,\L$ and $A_v$ stand for the quantum dimension and the quantum
$15j$ symbol \cite{cyk}. In 3d the $6j$ symbols replace the $15j$ symbols, and that state sum gives the Turaeev-Viro invariant \cite{tv}.

\bigskip
\bigskip
\noindent{\bf 2. The Barrett-Crane model}

\bigskip
\noindent Since GR is not a topological theory, the constraint (\ref{bcon}) has to be implemented, and therefore a different quantization route has to be followed. One can conjecture that exists a quantization procedure such that the quantities $B_\D$ become the 4d rotations algebra
operators $J_\D$, since the 4d rotation group irreps are labelling
the triangles $\D$ (or the dual faces $f$). Then one can show that
the constraint (\ref{bcon}) becomes a constraint on the triangle
irreps, given by
\be\e^{abcd}J_{ab}J_{cd}=0 \label{simp}\ee \cite{bce,bc}. In the
Euclidian case the irreps are given by the pairs of the $SU(2)$
spins $(j,j')$, so that the constraint (\ref{simp}) implies
$j=j'$. In the Minkowski case, requiring the hermiticity of the
$B$ operators implies that one needs the unitary irreps of the
Lorentz group. These are infinite-dimensional irreps and they are
given by the pairs $(j,p)$ where $j$ is the $SU(2)$ spin and $p$
is a continuous label. The constraint (\ref{simp}) implies that
$\L = (0,p)$ or $\L=(j,0)$.

One can argue that the spacelike triangles should be labelled by
the $(0,p)$ irreps, while the time-like triangles should be
labelled by the $(j,0)$ irreps. Since a spacetime triangulation
can be built from the spacelike triangles, Barrett and Crane have
proposed the following spin foam state sum (integral) for the
quantum general relativity \cite{bc}\be Z_{BC} = \int \prod_f p_f
dp_f \prod_v \tilde A_v (p_f) \quad, \label{bcss}\ee where $\tilde
A_v$ is an amplitude for the corresponding 4-simplex spin network,
given by \be \tilde A (p_1,\cdots,p_{10})= \int_{H^5}
\prod_{i=1}^5 dx_i \delta(x_1 -x_0) \prod_{i<j}K
_{p_{ij}}(x_i,x_j)\quad.\ee This is as an integral over the fifth
power of the hyperboloid $H=SO(3,1)/SO(3)$ of a propagator $K_p
(x,y)$ on that space. The propagator is given by \be  K_p (x,y)=
{\sin\left( p d(x,y)\right)\over p \sinh d(x,y)}\quad, \quad \cosh
d(x,y)= x\cdot y \quad.\ee

The expression (\ref{bcss}) is not finite for all triangulations,
but after a slight modification, consisting of including a
non-trivial edge amplitude $\tilde A (p_1,\cdots,p_4)$, the
partition function becomes finite for all non-degenerate
triangulations \cite{cpr}. This was a remarkable result, because
it gave a perturbatively finite quantum theory of gravity, which was
not based on string theory.

The main difficulties with the BC type models are:

1) The edge amplitudes are not determined in the BC approach, except by requiring the finiteness. By studying the converegence of the state sum, one can find choices with various degrees of convergence \cite{baecr}, but it is not clear which choice is the correct one. The reason why the edge amplitudes are not determined is that the BC quantization procedure is incomplete in the sense that it is not a direct quantization of a discretized path integral for GR, but one modifies a path integral for a topological gravity theory in order to implement the B constraint.

2) It is difficult to see what is the semi-classical limit, so that
it is not clear whether the corresponding effective action will be given by the EH
action plus the $O(l_P)$ corrections, where $l_P$ is the Planck length.

3) $Z_{BC}$ depends on a
triangulation, in accordance with the fact that 4d gravity is a
non-topological theory. However, the quantum gravity $Z$ must be a diffeomorphism invariant, and therefore it should be independent of the triangulation. One way to obtain such a $Z$ is to sum $Z_{BC}$ over the
triangulations, but this is difficult to do. Alternatively, one can try to define a continious limit of $Z_{BC}$ by taking increasingly finer triangulations, so that one would hopefully arrive at some effective diffeomorphism invariant action, in analogy to the 2d Ising model, where the discrete action at the critical point becomes a 2d diffeomorphism (conformally) invariant field theory action. 

4) Since the matter couples to the gravitational
field through the tetrads, one would need a formulation where the
basic fields are the tetrad one-forms instead of the composite $B$ 2-form. In the case of the
YM field, the coupling can be expressed in terms of the $B$ field
\cite{amm}, so that one can formulate a BC type models
\cite{amym,op}. However, for the fermions this is not possible, and
a tetrade based formulation is necessary. In \cite{amm} an
algebraic approach was proposed in order to avoid this problem,
and the idea was to use a result from the loop quantum gravity,
according to which the fermions appear as free ends of the spin
networks. Hence including open spin networks gives a new type of
spin foams \cite{amnsf}, and this opens a possibility of including
matter in the spin foam formalism. However, what is the precise
form of the matter spin foam amplitudes remains an open question.

\bigskip
\bigskip
\noindent{\bf{3. Spin foams for loop quantum gravity}}

\bigskip
\noindent One way to resolve the problems of the BC model is to use the spin foams in the loop quantum gravity formalism \cite{ro}. In \cite{amlqg} it was shown how to use the 3d spin foam state sum
invariants of embedded spin networks in order to define the physical states in the loop
quantum gravity formalism. The idea is to use the representation
of a quantum gravity state $|\Psi\rangle$ in the spin network
basis \be |\Psi\rangle = \sum_\g |\g\rangle\langle\g |\Psi\rangle
\quad.\ee The expansion coefficients are then invariants of the
embedded spin networks in the spatial manifold $\S$, and can be
formally expressed as \be \langle\g |\Psi\rangle = \int {\cal D} A\,
\langle\g |A\rangle\langle A|\Psi\rangle = \int {\cal D} A\, W_\g
[A]\,\Psi[A] \quad,\ee where $A$ is a 3d complex $SU(2)$
connection, $W_\g [A]$ is the spin network wave-functional
(generalization of the Wilson loop functional) and $\Psi[A]$ is a
holomorphic wave-functional satisfying the quantum gravity
constraints in the Ashtekar representation.

In the case of non-zero cosmological constant $\l$, a non-trivial
solution for $\Psi$ is known, i.e. the Kodama wavefunction \be \Psi[A] =
e^{\frac{1}{\l}\int_\S Tr\left(A\wedge dA +\frac23 A\wedge A\wedge
A\right)} \quad,\ee which is the exponent of the Chern-Simons action.
In the $\l=0$ case a class of formal
solutions is given by \be \Psi[A] = \prod_{x\in\S}\delta
(F_x)\,\Psi_0 [A] \quad,\ee i.e. a flat-connection wavefunction
\cite{amlqg}. In the $\l=0$ and $\Psi_0 = 1$ case the
corresponding spin network invariant is given by a 3d spin foam
state sum for the quantum $SU(2)$ at a root of unity \cite{amlqg}
\be \langle \g |\Psi\rangle = \sum_{j_f , \i_l} \prod_f \dim \,j_f \prod_v A_v (j_f, \i_l ,j_\g ,\i_\g ) \quad, \label{bfsni}\ee
where $A_v$ are the amplitides of the vertex spin networks. A vertex spin network is
given by the tetrahedron graph if no $\g$ vertex is present at the dual 2-complex vertex; otherwise it is given by a modified tetrahedron graph of a tetrahedron plus a spin network vertex connected by its edges to the tetrahedron verticies.   

In the $\l\ne 0$ case, the corresponding
spin network invariant is given in the Euclidian gravity case by
the Witten-Reshetikhi-Turaeev invariant for $q=e^{2\pi i/(k+2)}$,
where $k\in\bf N$ and $\l =k/l_P^2$, while in the Minkowski case, it is conjectured that
the invariant is given by an analytical continuation of the
Euclidian one, as $k\to ik$ \cite{smo}. Although there is no state sum representation of the WRT invariants, recently it has been shown that the square of the module of the WRT spin network invariants can be related through a linear transformation to the Turaeev-Viro spin network invariants\footnote{A Turaeev-Viro spin network invariant is defined as the TV state sum for a triangulation where a subset of the edges are marked by the irreps of a given spin network. On the other hand, the spin network invariant (\ref{bfsni}) is a state sum for a modified dual 2-complex, where a modification is obtained by inserting the spin network edges in the dual graph of a triangulation.} \cite{bar}.  

However, the problem with the Kodama and the $\d(F)$ wavefunctions is that they do not correspond to any particular value of the triads, so that these wavefunctions cannot describe the vacuum state of quantum gravity, which we define as a physical state which is peaked around the flat space triads $E_0$. In the $\l=0$ case, one can show that such a state is given by 
\be \Psi[A]= \d (F)\exp\left( i\int_\S d^3 x \,Tr(AE_0)\right)\ee
\cite{amv}, so that the corresponding spin network invariant is given by the state sum  
\be \langle \g |\Psi\rangle = \sum_{j_f , \i_l , j_l ,\i_v} \prod_f \dim \,j_f 
\prod_l C_{j_l}(E_0)\prod_v A_v (j_f, \i_l , j_l, \i_v ; j_\g ,\i_\g ) \quad, \ee
where 
\be C_{j}(E)= \int_{SU(2)} dg \,\bar \chi_j (g) \,e^{i\, Tr(AE)}\quad,\quad g=e^{iA}\quad,  \ee
and $A_v$ are the evaluations of modified tetrahedron spin networks. This modification comes from the fact that the dual two-complex is now labelled by two independent sets of irreps: $j_f$ label the faces, while $j_l$ label the edges\footnote{The intertwiner labels $\i_l$ and $\i_v$ are not independent labels. The $\i_l$ depends on the $j_f$'s meeting at the edge $l$ and $\i_v$ depends on the $j_l$'s meeting at the vertex $v$.}.

Given the invariants $I_\g = \langle \g |\Psi\rangle$ one can reconstruct the wavefunction in the triad representation as
\be \Phi [E] = \sum_\g I_\g \bar \Phi_\g [E] \quad,\label{wfe}\ee
where
\be \Phi_\g [E] = \int_{A\in \bf R} DA \exp\left( -i\int_\S d^3 x \,Tr(AE) \right) W_\g [A] \quad.\ee
This path integral can be defined by the state sum
\be \Phi_\g [E] = \sum_{j_l , \i_v ,\tilde\i_v } \prod_l C_{j_l}(E_l) \prod_v A_v (j_l ,\i_v ,j_\g , \i_\g , \tilde\i_v )\quad,\ee
where the vertex spin networks are obtained by composing a spin network $\G$ associated to the dual one-complex for a triangulation of $\S$ and the $\g$ spin network \cite{amv}.

In the $\l\ne 0$ case, one can argue that the modification of the wavefunction is given by $\Psi (A) = \Psi_K (A) \d (F - \l E_0)$ \cite{amv}, so that one would need to define the invariant
\be I_\g = \int DA \, e^{iS_{CS}[A]} \d (F - \l  E_0 ) W_\g [A] \quad. \ee
Once the functional $\Phi [E]$ is obtained, one can try to check the semiclassical limit by studying the effective equations of motion in the de-Broglie-Bohm formalism 
\be  \tilde p_i^a (E,\dot E, N ) = {\d S \over \d E_a^i} \quad,\quad S=Im(\log\Phi) \quad,\label{dbb}\ee
where $\tilde p_i^a$ is a canonically conjugate variable to the inverse triad density $E_a^i = (\det e ) e_a^i$, $\tilde p(E,\dot E ,N)$ is the expression for the $\tilde p$ in terms of the triad, its time derivative and the Lagrange multipliers $N$.

\newpage
\noindent{\bf{4. Tetrade spin foam model}}

\bigskip
\noindent The problem of coupling matter within the BC approach suggests that one should try to find a spin foam model which is based on the integration of the tetrad fields. 
This is feasible because the Palatini action is quadratic in the tetrads, so that the path integral over the tetrads is Gaussian and therefore one can write formally
\be Z=\int {\cal D} \o \, {\cal D} e \,e^{i\int \langle e^2 R \rangle} = \int {\cal D} \o \, (\det\, R )^{-1/2}\quad. \ee 
Hence one can try to define $Z$ as
\be Z= \int \prod_l dA_l \prod_f (\det \, F_f )^{-1/2}= \int \prod_l dg_l \prod_f 
w( g_f)\quad,\ee
where $\det \,F = (\e^{abcd}F_{ab}F_{cd})^2$, $g_f = e^{F_f}$ and $w(g_f ) =
(\det \, F_f )^{-1/2}$. Since
\be w(g) = \sum_\L c(\L )\,\chi_\L (g) \quad,\ee
where $c(\L)= \int_G dg \,\bar\chi_\L (g)\, w(g)$, we will obtain a state sum of the form
\be Z= \sum_{\L_f,\iota_l} \prod_f c(\L_f) \prod_v A_v
(\L_f,\iota_l) \quad.\label{grss}\ee

This state sum is of the same form as in the case of the topological theory; however, the weights we put on the faces are not $\dim \L_f$ but the functions $c(\L_f)$. It remains to be seen how the choice of the $c(\L_f)$ weights\footnote{These weights are given by the integrals which are generically divergent, due to $\det\, F_f  = 0$ configurations, so that some kind of regularization must be used.} will affect the convergence of the partition function $Z$, and whether or not one would need to use the quantum group in order to achieve the finiteness of the $Z$. 

As far as the coupling of matter is concerned, as well as including the cosmological constant term, this would require the evaluation
of the partition function with the sources (generating functional)
\be  Z (J,j)=\int {\cal D} \o \, {\cal D} e \,e^{i\int \langle e^2 R \rangle+ Tr(J\o ) + Tr(je) } \quad,\ee
which can be formally rewritten as
\be Z (J,j) = \int {\cal D} \o \, e^{i\int Tr(J\o )} (\det R )^{-1/2} e^{-i\int \langle j R^{-1}j \rangle /4}\quad. \ee
This expression can be defined on a spacetime triangulation along the lines of the $J=j=0$ case, but when the sources are present a more intricate state sum will appear. It can be defined as
\be Z= \int \prod_l dg_l \,u(g_l ,J_l )\prod_f w(g_f ,j_\e ) \quad,\ee
where
\be u(g_l ,J_l ) = e^{iTr(\o_l J_l )} \quad,\quad w(g_f ,j_\e )= w(g_f) e^{-i \langle j_\e R_f^{-1} j_{\tilde\e}\rangle/4 } \quad,\ee
and $\e$, $\tilde\e$ are two different edges of the triangle $\D$ dual to the face $f$. By expanding the functions $u$ and $w$ as
\be u = \sum_{\L_l} \a (\L_l ,J_l) \chi_{\L_l }(g_l) \quad,\quad w = \sum_{\L_f} \b (\L_f ,j_\e) \chi_{\L_f }(g_f)\quad,\ee
and performing the group integrations, one obtains a state sum
\be Z(J,j)=\sum_{\L_f , \i_l , \L_l ,\i_v} \prod_f \b (\L_f , j_\e) 
\prod_l \a (\L_l ,J_l) \prod_v A_v (\L_f, \i_l , \L_l, \i_v ) \quad,\ee
which similarly to the sum (\ref{bfsni}) involves a dual 2-complex whose edges and faces are independently colored with the irreps of the relevant group ($SO(4)$ in the Euclidian gravity case, or $SO(3,1)$ in the Minkowski case).

\bigskip
\bigskip
\noindent{\bf{5. Conclusions}}

\bigskip
\noindent The two spin foam models we have described have an advantage over the BC type models in the fact that they have been formulated as direct quantizations of GR, so that all the simplex amplitudes are fixed. Also these are the models where the matter can be more easilly introduced. As far as the semiclassical/continious limit is concerned, they seem to be promising candidates. In the loop quantum gravity case, the model is defined in the continuum space, and 
the problem is in finding an aproximation for the sum (\ref{wfe}) such that the equations (\ref{dbb}) give
Planck length corrected Einstein equations. In the tetrade model, it appears easier to extract the semiclassical limit by considering finer (larger) triangulations then summing over the triangulations, but a technique must be developed in order to do this, perhaps something analogous to the 2d Ising model near the critical point. Clearly, much more work is necessary in order to resolve these issues.

On the mathematical side, the expression (\ref{bfsni}) is a new way to calculate the knot and spin network invariants, which can be easilly generalised to higher dimensions and it is more efficient way of calculating these invariants than the TV state sum approach \cite{amstr}.

\end{document}